\documentclass[pra,aps,groupedaddress,twocolumn,floatfix,nofootinbib]{revtex4-2}

\usepackage{graphicx}
\newcommand{\beq}{\begin{equation}}
\newcommand{\eeq}{\end{equation}}
\newcommand{\beqa}{\begin{eqnarray}}
\newcommand{\eeqa}{\end{eqnarray}}

\def\opone{\leavevmode\hbox{\small1\normalsize\kern-.33em1}}

\begin{document}

\title{Quantum non-locality:\\ from denigration to the Nobel prize, via quantum cryptography}

\author{Nicolas Gisin}
\affiliation{Group of Applied Physics, University of Geneva, 1211 Geneva 4, Switzerland\\
Constructor University, Geneva, Switzerland}

\date{\small \today}

\maketitle

In the late 1960s, a young physicist was sailing along the coast of California towards Berkeley, where he got a post-doc position in astronomy. But his real goal was not astronomy, at least not immediately. First, John Clauser eagerly wanted to test some predictions of quantum theory that were at odds with a then recent and mostly ignored result by an Irish physicist John Stewart Bell, working at the celebrated CERN near Geneva. Bell, inspired by David Bohm's hidden variable model of quantum theory, proved that all possible correlations that can be described by {\it local} variables necessarily satisfy some inequalities, today known as Bell inequalities. These inequalities are mathematically quite trivial. However, quantum theory predicts that they can be violated even when the correlation is between outcomes of far distant measurements. Denote $a$ and $b$ the measurement outcomes and $x$ and $y$ the measurement settings (e.g. polarizers' orientations), and denote $\lambda$ the hypothetical hidden local variables. Accordingly, the entire statistics of the experiment is captured by the so-called ``correlation" - strictly speaking, conditional probability distribution - $p(a,b|x,y,\lambda)$. The $\lambda$'s are hidden in the sense that they are not part of quantum theory, though the usual quantum state $\psi$ could well be a part of $\lambda$. Here, local - or Bell-local - refers to the assumption that the correlation factorizes in two parts, one for each of the distant sides of the experiment:
\beq\label{BellLocal}
p(a,b|x,y,\lambda)=p(a|x,\lambda)\cdot p(b|y,\lambda)
\eeq
That's the only assumption necessary to derive Bell inequalities. The $\lambda$'s denote the state of the system as described by any possible future physical theory (except that the settings $x$ and $y$ are assumed to be independent of $\lambda$). In this sense, Bell inequalities go way beyond quantum theory: a violation of a Bell inequality proves that no future theory can satisfy the locality condition (\ref{BellLocal}).

John Clauser, Abner Shimony, Michael Horne and Richard Holt were among the very few who understood this in the 1960s and all wanted to test Bell inequalities, Clauser to prove quantum theory wrong, Holt, a young student at Harvard, to prove the Bell-locality assumption (\ref{BellLocal}) wrong. Clauser was in a good position thanks to existing equipment at Berkeley. Indeed, Carl Kocher had done a similar experiment in 1967, though for other purposes. Unfortunately, Kocher, and even earlier Chien-Shiung Wu, had only measured the correlations when the polarizers were either parallel or orthogonal, while a proper violation of Bell inequality requires intermediate orientations. Note that assuming that polarization is a 2-dimensional quantum system, a qubit as one says today, correlations at $45^o$ can be derived from the parallel and orthogonal correlations assuming no-signaling \cite{GisinHPA}: $E_{45}=(E_{\parallel}+E_{\perp})/\sqrt{2}$. That wasn't known at the time. But regardless, the visibilities measured by Kocher and Wu were below 50\%, while a proper violation requires visibilities larger than 71\%. Hence the race was on. Clauser got there first, confirming quantum predictions, against his expectation. But then Holt obtained his own result, confirming the inequality, against his expectation. Somehow, the score was one to one.

At that time, these fascinating and intriguing results interested almost no one, except some hippies who could later claim to have saved physics \cite{Hippies}. Clauser had long discussions with them, though the last time I met him he had turned into a loud climate skeptic.

In the 1970's, my friend Alain Aspect was doing his French civil service in Africa, reading physics, as we all do. When he hit on Bell inequalities, it was love at first sight: ``I want to work on that".
Back in Paris, he traveled to Geneva to meet John Bell and told him about his plans. Bell replied: ``Do you have a permanent position?". Indeed, in those times, working on - or even just showing interest in - Bell inequalities was a kind of scientific suicide. Bohr had it all solved, went the dogma. Looking back, it is difficult to appreciate how deeply denigrated was all research around Bell inequalities and entanglement - the quantum resource necessary to violate them. At the time, French had no agreed-upon translation of entanglement, some used ``enchev\^etrement", others ``intrication" (the latter has by now been officially recognized by the French academy). 

Fortunately, the French system allowed young physicists such as Aspect to hold permanent positions, so he decided to score the winning goal. Crucially, he planned to add fast switches that would allow one to chose the measurement settings $x$ and $y$ while the photons were already too far away to possibly influence the other side. 
Aspect was able to achieve this using newly developed lasers to pump his entanglement source, while Clauser and Holt had to use flash lamps. In a series of three beautiful experiments in the early 1980's, Aspect settled the dispute in favor of quantum theory. Accordingly, no future theory will ever satisfy the locality condition (\ref{BellLocal}). Today, this is often expressed by the short expression non-local, which really means not-Bell-local.

Despite these beautiful experiments and the intellectually fascinating discoveries, Bell inequalities remained dismissed and poorly understood. Even to this day, the clear terminology non-local (equivalently, not-Bell-local) is too often blurred as not satisfying ``local-realism", as if non-realism was a way out \cite{GisinNonRealism,Laudisa}. The fact is that assumption (\ref{BellLocal}) is no longer tenable. 
As an example, consider the scientific background provided by the Nobel Committee \cite{{NobelScBackground}}. A few lines after correctly presenting Bohm's non-local hidden variable model, one reads that Bell inequality violation shows ``that no hidden variable theory would be able to reproduce all the results of quantum mechanics", contradicting the just cited Bohm model (which does predict violation of Bell inequalities). The correct statement is that no \emph{local} variable theory is able to reproduce all results of quantum mechanics. And a few lines further, locality is defined as no-signaling - no communication without any physical object carrying the information, despite the fact that one of the main contribution of quantum information to the foundations of physics is a clear distinction between these two concepts. Next, realism is defined as determinism, even though Bell inequalities also hold in all stochastic theories satisfying (\ref{BellLocal}). All this illustrates that Bell inequalities are still poorly understood by the general physics community. The 2022 Nobel Prize in physics allows one to hope that henceforth Bell inequalities will
be part of all physics cursus.

One major step towards a better appreciation of Bell inequalities came from a young Polish PhD student at Oxford University, Artur Ekert. In 1991, he realized that non-local quantum correlations are nothing but cryptographic keys! Indeed, in both cases, the correlation is private and, after some error correction, the bits on both sides are identical. This proposal to exploit non-local correlations for cryptographic applications changed everything (though it took several years to prove Ekert's intuition correct \cite{DIQKD,RMP-NL-Brunner14}). Moreover, just a few years later, Peter Shor showed how one can exploit entanglement to break the commonly used public key cryptography system RSA. Thus, in the 1990's, non-locality and entanglement were in the spotlight, at last.

But that would not have sufficed. The entanglement source used so far was too complex. Leonard Mandel, at Rochester University, realized that a humble non-linear crystal could provide highly entangled photons when pumped by a simple diode. Moreover, the entangled photons could easily be coupled into optical fibers, opening thus the road to quantum cryptography using existing infrastructure, e.g. our demonstrations of quantum non-locality over the Swisscom network and quantum cryptography under Lake Geneva \cite{QKQLakeGeneva}, illustrated in Fig. 1. 

\begin{figure}[h]
\includegraphics[width=8cm]{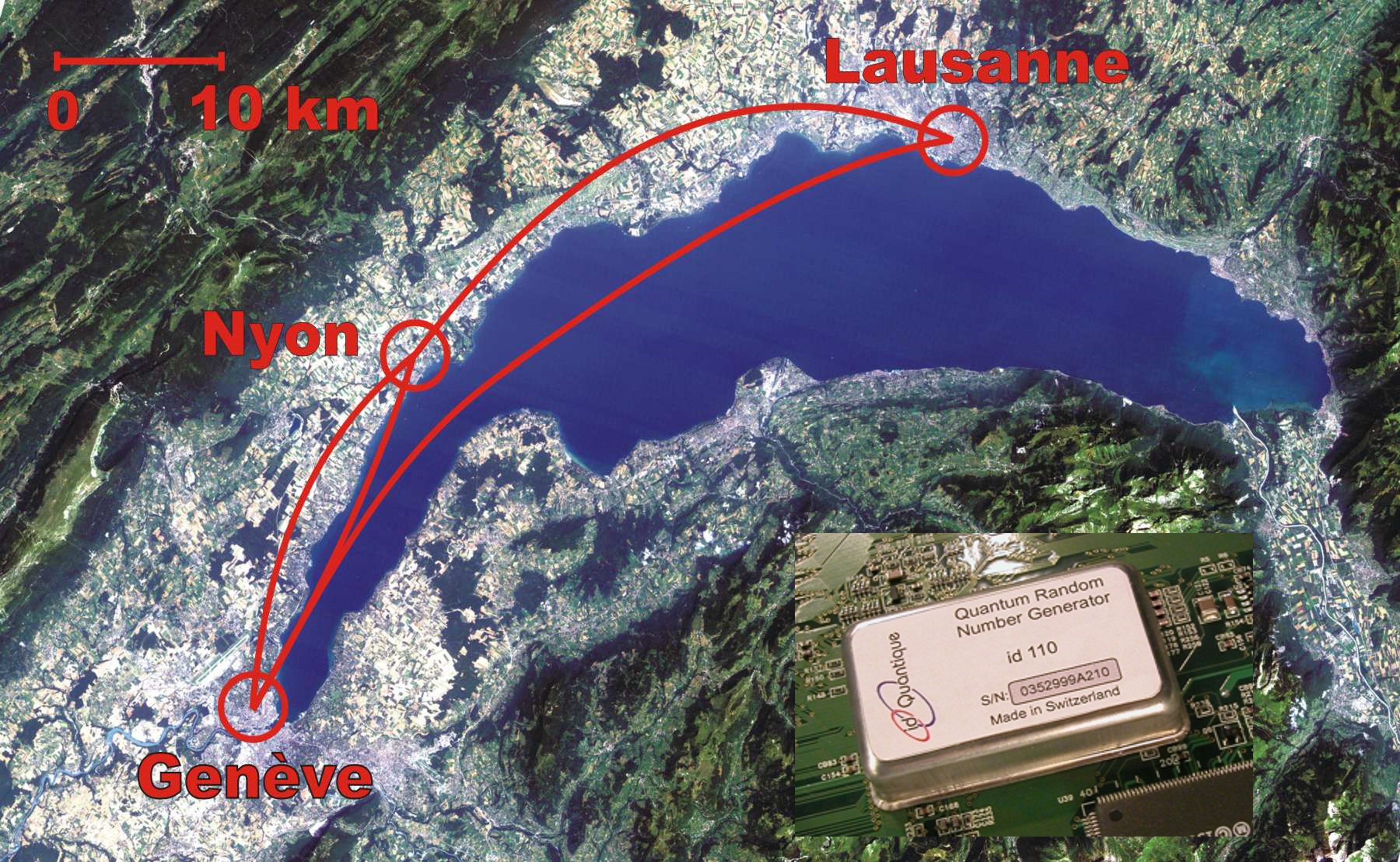}
\caption{\it Quantum cryptography under Lake Geneva was the first quantum experiment requiring a satellite photo to illustrate it. Nowadays in commercial use \cite{QKD-IDQ}.}
\end{figure} 

The focus thus changed from foundations of quantum physics to quantum information science and technologies. New ideas emerged, like quantum teleportation and quantum error correction, in addition to fast experimental progress. Anton Zeilinger had been interested in foundations since his early days as a physicist in neutron interferometry. He quickly joined the quantum information community and became a leading figure. His demonstration of quantum teleportation, immediately after the one in Rome, by De Martini and Popescu, attracted enormous attention, both within the scientific community and from the public at large. Soon thereafter, Zeilinger went further and demonstrated the teleportation of entanglement. 
Generally, quantum teleportation is the resource behind quantum computation and long-distance quantum communication. Zeilinger also improved on Aspect's experiments in fast choices of the measurement settings (though with low detection efficiencies, hence much simpler experiments were possible \cite{ZbindenSwitch}). Next, the so-called detection loophole was closed in an optical experiment by Paul Kwiat's group at Ilinois University \cite{DetLoopKwiat}, before a series of ``final" loophole-free experiments, one of which was carried out by Zeilinger's group. Zeilinger continued with a long series of remarkable experiments, including dense coding, and the demonstration of 3 and 4 photon entanglement, culminating with the long-distance free-space communication in the Canary Islands, making him a clear leader of the new field of experimental quantum information.

The next step was the understanding that entanglement is actually not necessary for point-to-point quantum cryptography, though it remains essential for the security proofs. Indeed, while in tests of Bell inequalities one sets the source about half way between the measurement devices, in applications it is much more practical to put it on one side. Hence, a mere single photon travels to the receiver. This, and a few additional tricks - especially the move to wavelengths compatible with standard telecom optical fibers, which we initiated in Geneva with the development of specific single-photon detectors, made it possible to not only demonstrate quantum cryptography, but to industrialize and commercialize it. 
Today, there are many small companies selling quantum cryptography equipments, some quite advanced as illustrated in Fig. 2. Development efforts will continue, but no longer with the aim of excluding local variables satisfying (\ref{BellLocal}); the goals now are to make the equipments cheaper, faster and able to cover longer distances, probably by exploiting quantum teleportation.

On the conceptual side, the violation of Bell inequalities dramatically revolutionized our world-view. Interestingly, Newton's theory of gravity was also non-local, even signaling. But Einstein improved on it, making gravity local. It is thus not surprising that he strongly objected to quantum non-locality, not fully appreciating that it is of a very different sort: without any action at a distance, just non-local randomness without any possibility to use it for signaling \cite{Gilder,Qchance}. In contrast to Newton's non-locality, quantum non-locality is here to stay; the experimental evidence is clear on that point. Today, non-local quantum correlations are explored for device-independent quantum information processing \cite{RMP-NL-Brunner14}, in particular device-independent quantum cryptography, a truly fascinating research field unthinkable before Bell's work. Another timely and exciting conceptual goal is to take non-locality beyond the simple Bell scenario and place it in the context of quantum networks with several independent sources of entanglement \cite{NLN}; this already led to the remarkable result that some quantum networks can't be described using only real-number Hilbert spaces \cite{realH}.

Sincere congratulations John, Alain and Anton, you made me so happy. Congratulations also to the Nobel Committee for recognizing, finally, the game-changing findings of the late John Stewart Bell, with whom, along with his wife Mary, I had the pleasure of sharing several cheese raclettes in downtown Geneva.

\begin{figure}[h]
\includegraphics[width=9cm]{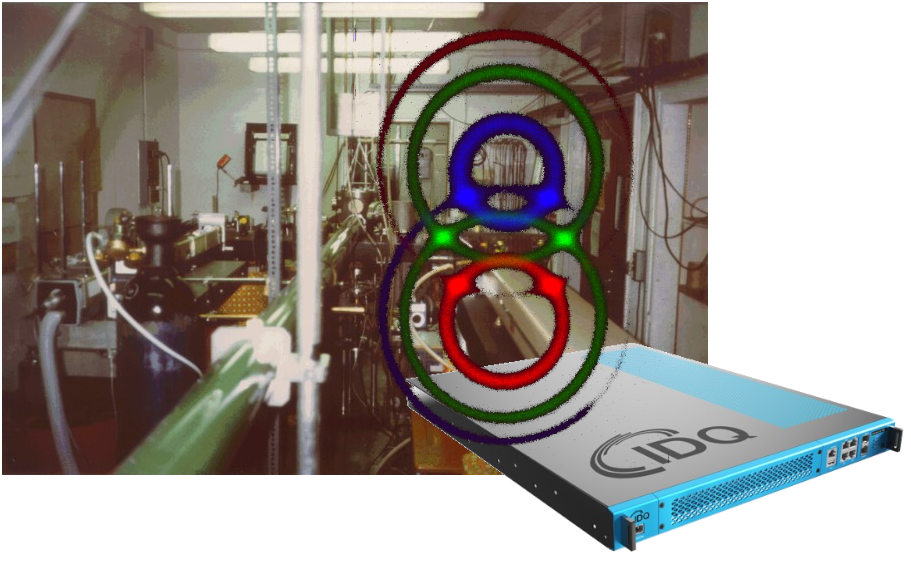}
\caption{\it A picture of Aspect's lab around 1981, followed by a commercial quantum cryptography equipment \cite{QKD-IDQ}, with the entanglement source used by Zeilinger in the middle (photo by Paul Kwiat and Michael Reck).}
\end{figure}

\small{
\section*{Acknowledgement}
Many thanks to Benjamin Feddersen for polishing my English.}

\end{document}